\documentclass{article}
\date{}

\usepackage[utf8]{inputenc}

\usepackage{graphicx}
\usepackage{cprotect}

\usepackage{jcappub}

\usepackage{listings}
\usepackage[dvipsnames]{xcolor}

\definecolor{halfgray}{gray}{0.55}
\definecolor{ipython_frame}{RGB}{207, 207, 207}
\definecolor{ipython_bg}{RGB}{247, 247, 247}
\definecolor{ipython_red}{RGB}{186, 33, 33}
\definecolor{ipython_green}{RGB}{0, 128, 0}
\definecolor{ipython_cyan}{RGB}{64, 128, 128}
\definecolor{ipython_purple}{RGB}{170, 34, 255}

\usepackage{listings}
\lstdefinelanguage{iPython}{
    morekeywords={access,and,break,class,continue,def,del,elif,else,except,exec,finally,for,from,global,if,import,in,is,lambda,not,or,pass,print,raise,return,try,while},%
    morekeywords=[2]{abs,all,any,basestring,bin,bool,bytearray,callable,chr,classmethod,cmp,compile,complex,delattr,dict,dir,divmod,enumerate,eval,execfile,file,filter,float,format,frozenset,getattr,globals,hasattr,hash,help,hex,id,input,int,isinstance,issubclass,iter,len,list,locals,long,map,max,memoryview,min,next,object,oct,open,ord,pow,property,range,raw_input,reduce,reload,repr,reversed,round,set,setattr,slice,sorted,staticmethod,str,sum,super,tuple,type,unichr,unicode,vars,xrange,zip,apply,buffer,coerce,intern},%
    sensitive=true,%
    morecomment=[l]\#,%
    morestring=[b]',%
    morestring=[b]",%
    morestring=[s]{'''}{'''},
    morestring=[s]{"""}{"""},
    morestring=[s]{r'}{'},
    morestring=[s]{r"}{"},%
    morestring=[s]{r'''}{'''},%
    morestring=[s]{r"""}{"""},%
    morestring=[s]{u'}{'},
    morestring=[s]{u"}{"},%
    morestring=[s]{u'''}{'''},%
    morestring=[s]{u"""}{"""},%
    literate=
    *{+}{{{\color{ipython_purple}+}}}1
    {-}{{{\color{ipython_purple}-}}}1
    {*}{{{\color{ipython_purple}$^\ast$}}}1
    {/}{{{\color{ipython_purple}/}}}1
    {^}{{{\color{ipython_purple}\^{}}}}1
    {?}{{{\color{ipython_purple}?}}}1
    {!}{{{\color{ipython_purple}!}}}1
    {\%}{{{\color{ipython_purple}\%}}}1
    {<}{{{\color{ipython_purple}<}}}1
    {>}{{{\color{ipython_purple}>}}}1
    {|}{{{\color{ipython_purple}|}}}1
    {\&}{{{\color{ipython_purple}\&}}}1
    {~}{{{\color{ipython_purple}~}}}1
    {==}{{{\color{ipython_purple}==}}}2
    {<=}{{{\color{ipython_purple}<=}}}2
    {>=}{{{\color{ipython_purple}>=}}}2
    {+=}{{{+=}}}2
    {-=}{{{-=}}}2
    {*=}{{{$^\ast$=}}}2
    {/=}{{{/=}}}2,
    literate=
    {á}{{\'a}}1 {é}{{\'e}}1 {í}{{\'i}}1 {ó}{{\'o}}1 {ú}{{\'u}}1
    {Á}{{\'A}}1 {É}{{\'E}}1 {Í}{{\'I}}1 {Ó}{{\'O}}1 {Ú}{{\'U}}1
    {à}{{\`a}}1 {è}{{\`e}}1 {ì}{{\`i}}1 {ò}{{\`o}}1 {ù}{{\`u}}1
    {À}{{\`A}}1 {È}{{\'E}}1 {Ì}{{\`I}}1 {Ò}{{\`O}}1 {Ù}{{\`U}}1
    {ä}{{\"a}}1 {ë}{{\"e}}1 {ï}{{\"i}}1 {ö}{{\"o}}1 {ü}{{\"u}}1
    {Ä}{{\"A}}1 {Ë}{{\"E}}1 {Ï}{{\"I}}1 {Ö}{{\"O}}1 {Ü}{{\"U}}1
    {â}{{\^a}}1 {ê}{{\^e}}1 {î}{{\^i}}1 {ô}{{\^o}}1 {û}{{\^u}}1
    {Â}{{\^A}}1 {Ê}{{\^E}}1 {Î}{{\^I}}1 {Ô}{{\^O}}1 {Û}{{\^U}}1
    {œ}{{\oe}}1 {Œ}{{\OE}}1 {æ}{{\ae}}1 {Æ}{{\AE}}1 {ß}{{\ss}}1
    {ç}{{\c c}}1 {Ç}{{\c C}}1 {ø}{{\o}}1 {å}{{\r a}}1 {Å}{{\r A}}1
    {€}{{\EUR}}1 {£}{{\pounds}}1,
    commentstyle=\color{ipython_cyan}\ttfamily,
    stringstyle=\color{ipython_red}\ttfamily,
    keepspaces=true,
    showspaces=false,
    showstringspaces=false,
    rulecolor=\color{ipython_frame},
    frame=single,
    frameround={t}{t}{t}{t},
    numberstyle=\tiny\color{halfgray},
    backgroundcolor=\color{ipython_bg},
    basicstyle=\scriptsize\ttfamily,
    keywordstyle=\color{ipython_green}\ttfamily,
}

\usepackage{amsmath, amssymb}
\usepackage{diagbox}
\usepackage{comment}

\usepackage[normalem]{ulem}
\def\d{{\rm d}}
\def\dNdS{{\rm d}N/{\rm d}S}
\newcommand{\Fermi}{{\it Fermi}}
\newcommand{\g}{$\gamma$}

\usepackage{hyperref}

\title{Deepening gamma-ray point-source catalogues with sub-threshold information}
\author[a, 1]{Aurelio Amerio,\note{Corresponding author.}}
\author[b]{Francesca Calore,}
\author[b]{Pasquale Dario Serpico,}
\author[a]{Bryan Zaldivar}

\affiliation[a]{Instituto de F\'isica Corpuscular (IFIC), University of Valencia and CSIC, Calle Catedrático José Beltrán 2, 46980 Paterna, Spain}
\affiliation[b]{Laboratoire d'Annecy de Physique Th\'eorique (LAPTh), CNRS, Univ. Savoie Mont Blanc, F-74940 Annecy, France}

\emailAdd{aurelio.amerio@ific.uv.es}
\emailAdd{calore@lapth.cnrs.fr}
\emailAdd{serpico@lapth.cnrs.fr}
\emailAdd{b.zaldivar.m@csic.es}

\abstract{
We propose a novel statistical method to extend \Fermi-LAT catalogues of high-latitude \g-ray sources below their nominal threshold. 
To do so, we rely on the determination of the differential source-count distribution of sub-threshold sources which only provides the statistical flux distribution of faint sources. 
By simulating ensembles of synthetic skies, we assess quantitatively the likelihood for pixels in the sky with relatively low-test statistics to be due to sources, therefore complementing the source-count distribution with spatial information.
Besides being useful to orient efforts towards multi-messenger and multi-wavelength  identification of new \g-ray sources, we expect the results to be especially advantageous for statistical applications such as cross-correlation analyses. 
}

\begin{document}
\begin{flushright}
LAPTH-035/23
\end{flushright}

\maketitle

\section{Introduction}

Our view of the high-energy $\gamma$-ray sky has been revolutionised  by the Large Area Telescope (LAT) onboard the \Fermi~satellite, which is on its surveying mission since 2008: Since the publication of the fourth source  catalogue (4FGL) based on 8 years of data~\cite{Fermi-LAT:2019yla}, incremental updates appear periodically. The latest incarnation, the data release 3 (DR3) based on 12 yrs of data~\cite{Fermi-LAT:2022byn}, includes 6658 point-like sources in the energy range from 50 MeV to 1 TeV, with extragalactic blazars constituting the largest associated class. Apart for revealing entirely new classes of objects (such as Galactic millisecond pulsars~\cite{Caraveo:2013lra}), the explosion of the number of $\gamma$-ray sources has allowed for numerous applications in multi-wavelength and multi-messenger astrophysics and astroparticle physics. Unsurprisingly, the essential requirement for a source to enter a catalogue is  that its signal strength is significantly above the background, dominated by $\gamma$ rays associated to energy-loss processes of cosmic rays in the interstellar gas and radiation field. Since the pioneering analysis of EGRET data~\cite{Mattox:1996zz}, the signal strength is typically quantified by a test statistics $2\ln({\cal L}/{\cal L}_0)$,  comparing the maximum value of the likelihood function including the source, ${\cal L}$, vs. the one without the source, ${\cal L}_0$. 

However, many more sources are believed to hide below the detection threshold. In particular, a significant part of the quasi-isotropic, high Galactic latitude emission, is attributed to point-like sources too dim to be individually detected by the LAT~\cite{Fornasa:2015qua}. Nonetheless, a statistical approach relying on the pixel-count distribution has long been suggested to provide a diagnostic tool to 
separate truly diffuse $\gamma$-ray signals from unresolved sources roughly isotropically distributed~\cite{Dodelson:2009ih}. Since its pioneering application in~\cite{Malyshev:2011zi}, a number of articles have used it~\cite{Zechlin:2015wdz,Zechlin:2016pme,Lisanti:2016jub}
to extend the measurement of the differential source-count, i.e.~$\d N/\d S$, below the detection threshold of \Fermi-LAT catalogues, as well as to probe extragalactic source populations~\cite{Manconi:2019ynl} and dark matter properties~\cite{Zechlin:2017uzo,Chang:2018bpt}.

Recently, in~\cite{Amerio:2023uet} one of us extended these results by adopting machine learning techniques,  obtaining  a $\d N/\d S$ (number of sources per unit flux) distribution which  is in excellent agreement in the resolved regime with the one derived from catalogues, while extending as $\d N/\d S\propto S^{-2}$ in the unresolved regime down to fluxes of about $5\times 10^{-12}\,$ cm$^{-2}$ s$^{-1}$. 
In~\cite{Amerio:2023uet}, the $\dNdS$ has been deduced by training a neural network to reconstruct the source-count distribution of the high-latitude sky given a set of simulations with broad priors for the $\dNdS$ parameters. The core idea was that if a neural network can learn how to correctly reconstruct the $\dNdS$ for a series of simulated sky maps whose prior is broad enough, and if the simulator is accurate enough, then the neural network can determine correctly the source-count distribution function given the \Fermi~photon counts map. 
The output of~\cite{Amerio:2023uet}, i.e.~$\dNdS$, describes the number of gamma-ray sources per differential unit of flux. This quantity can only be statistically used to describe the unresolved gamma-ray source population, and gives us no information concerning the location of any of the sources.

The goal of the present work is instead to probabilistically extend the \Fermi-LAT source catalogue exploiting the output of~\cite{Amerio:2023uet}. We will do so by proposing a new methodology applied to the \Fermi-LAT data which consists in a) proposing a test statistic (TS) for the gamma-ray counts in the sky, whose distributions across spatial pixels corresponding to the \Fermi-LAT measured map and those coming from the simulations can be compared; and b) identifying the pixels in the measured map whose TS values are large enough depending on the previous comparison, therefore overcoming the limitation of~\cite{Amerio:2023uet} of not being able to provide spatial information.

The new methodological development we propose in this work can be exploited to obtain an extended catalogue, final product of this analysis, which, albeit only probabilistically defined, can be nonetheless very useful, for instance, in multi-wavelength (e.g.~\cite{2015ApJS..217....2M,Pena-Herazo:2021yrp,Pena-Herazo:2021ybj}) or in cross-correlation~\cite{2009ApJ...707L..56K,2010MNRAS.407..791G,Turley:2018biv,Li:2022vsb} analyses.   
Other methods to build $\gamma$-ray probabilistic catalogues have been attempted in the past, e.g.~\cite{Daylan:2016tia}, which however has been applied to limited regions of the real sky.

This article is structured as follows: In Sec.~\ref{sec:data} we describe our data selection and introduce the map-making models we  use in our analysis.  Sec.~\ref {sec:procedure} is devoted to the quantitative setting of the problem and the statistical procedure followed. In Sec.~\ref{sec:catalogue} we present our results, which are also  made publicly available at \href{https://doi.org/10.5281/zenodo.8070852}{https://doi.org/10.5281/zenodo.8070852}. In Sec.~\ref{sec:conclusions} we outline some perspectives and present our conclusions.  

\section{Data selection and model components}
\label{sec:data}
\subsection{Data selection}
We consider the updated 14 years \Fermi-LAT data set (from week 9 to week 745) in the $(1,10)$ GeV energy range~\footnote{The energy range is chosen for consistency with the $\dNdS$ determination used below.} and Pass 8 event selection \cite{Fermi-LAT:2013jgq,Bruel:2018lac}. This ensures a good balance between high statistics and good angular resolution of the detector. We employ the \texttt{Fermi Science Tools} suite version 2.0.8 ~\cite{Fermitools} to analyse \Fermi-LAT data with the following settings: 
We adopt the \verb|P8R3_SOURCEVETO_V3| instrument response functions (IRF), event class (EVCLASS) 2048 (\texttt{SOURCEVETO}) and event type (EVTYPE) 1 (\verb|FRONT|). We used standard quality selection criteria, i.e. \verb|DATA QUAL==1| and \verb|LAT CONFIG==1|.
Atmospheric \g~rays from the Earth limb emission are removed by adopting a cut on the maximum zenith angle \texttt{ZMAX} of 90 degrees. We consider \verb|FRONT| events in order to have an optimal angular resolution, and the \texttt{SOURCEVETO} class of events, in order to have a good suppression of the charged cosmic-ray background while still retaining a large event statistics. 
In table~\ref{tab:settings}, we provide the reader with  a summary of the settings.

We employ the \texttt{Fermi Science Tools} to compute photon counts map, exposure map and point spread function (PSF) of the LAT.
The maps are binned in $N_{\rm pix}$ pixels with equal area  relying on \texttt{HEALPix}.~\footnote{\href{https://healpix.sourceforge.io}{https://healpix.sourceforge.io}} The chosen pixelization is expressed in terms of the ``resolution parameter'' $N_{\rm side}$ controlling the number of subdivisions of a great circle on the sphere, related to $N_{\rm pix}$ via $N_{\rm pix}=12 \, N_{\rm side}^2$. 
As we  motivate in section~\ref{sec:catalogue}, we simulate the sky
with $N_{\rm side}$ = 1024, but run the pixel analysis with $N_{\rm side}$ = 512. 

We perform an energy-integrated analysis in the $(1,10)$ GeV energy range, but our procedure can be  extended to multiple energy bins.

\begin{table}[t]
\centering
\begin{tabular}{ | l| l|  }
\hline
Healpix order & 9, 10  \\ 
Weeks & $9-745$  \\ 
Emin & 1 GeV \\
Emax & 10 GeV \\
Instrument Response Functions (IRFs) & \verb|P8R3_SOURCEVETO_V3| \\
EVCLASS & 2048 (Source Veto)  \\ 
EVTYPE & 1 (Front)  \\
ZMAX & 90 \\
\hline
\end{tabular}
\caption{\texttt{Fermi Science Tools} settings used for the 14-year data set analysis. }
\label{tab:settings}
\end{table}

\subsection{Null- and alternative-hypothesis models}

Our goal is to devise an algorithm able to detect point sources as \g-ray excesses
on top of a Poisson-distributed model for the background diffuse emission components. 
In this respect, our null-hypothesis model only consists of diffuse emission components, while
the alternative hypothesis model also contains point sources. 

We construct our background-only map $\mathcal{B}$ including two components:
\begin{equation}
    \mathcal{B} =  A_{\rm gal}\cdot\mathcal{G} + F_{0}\,,
    \label{eq:mapB}
\end{equation}
where 
\begin{itemize}
    \item $F_{0}$ is an isotropic (usually assumed extragalactic) component accounting for the average flux of all the unresolved  sources as well as possible diffuse emission mechanisms. 
    \item $\mathcal{G}$ is  the \verb|gll_iem_v07| template morphology of the diffuse \g-ray emission from the Milky Way provided by the \Fermi-LAT collaboration~\footnote{\href{https://fermi.gsfc.nasa.gov/ssc/data/access/lat/BackgroundModels.html}{https://fermi.gsfc.nasa.gov/ssc/data/access/lat/BackgroundModels.html}}, with normalization $A_{\rm gal}$.      
\end{itemize}
The parameters $A_{\rm gal}$ and $F_0$ are determined via a fit to the \Fermi-LAT photon counts map, as explained in section~\ref{sec:procedure}. In order to obtain a background map in units of counts in each of the $N_{\rm pix}$ pixels, we multiply by the exposure map $\mathcal{E}$ and the steradian-to-pixel conversion factor $4\pi / N_{\rm pix}$:
\begin{equation}
    C_\mathcal{B} = \frac{4\pi}{N_{\rm pix}} \int_{1 \, \rm GeV}^{10 \, \rm GeV} \mathcal{B}(E)\, \mathcal{E}(E){\rm d}E \,.
\end{equation}
The exposure map is extracted using the \texttt{Fermi Science Tools} for 10 logarithmic bins in the (1, 10) GeV range (and treated as the corresponding piecewise function when performing the integral), in order to be consistent with the binning of the Galactic foreground model.
We also account for the PSF, responsible for an angular smoothing of the map,  via the \texttt{Fermi Science Tools}. 
Similarly to the map ${\cal E}$, we average over the (1, 10) GeV range  with a $E^{-2.4}$ weight, according to the overall energy dependence of the data  at high Galactic latitudes \cite{Fornasa:2016ohl}.

\medskip 
On the other hand, our first alternative hypothesis model is based on the 4FGL-DR3, and will be used to validate our procedure and TS analysis on the official \Fermi~catalogue. 
We generate a new map 
\begin{equation}
    \mathcal{K} = {\cal C}+ A_{\rm gal}\cdot\mathcal{G} + F_{0}\,,
    \label{eq:mapK}
\end{equation}
including the same background model as in eq.~\eqref{eq:mapB} (i.e.~same parameters for $A_{\rm gal}$ and $F_0$), and adding a synthetic map associated to the list of sources in the 4FGL-DR3 catalogue with $|b|\geq 30^\circ$, ${\cal C}$. The map is appropriately smoothed via the PSF and convoluted with the exposure map, and converted into a pixelized count map, as previously described. 

\medskip
Finally,  as our second alternative hypothesis we consider a model for point sources based on the source-count distribution 
derived in~\cite{Amerio:2023uet}.
Analogously to the catalogue, we include an additional source component, $\mathcal{S}$, on top of the background-only model:
\begin{equation}
    \mathcal{M} = \mathcal{S} + A_{\rm gal}\cdot\mathcal{G} + F_{\rm  iso}
    \label{eq:map}
\end{equation}
The map $\mathcal{S}$ is constituted of point sources randomly placed  on the sphere, drawn according to the $\dNdS$ inferred in~\cite{Amerio:2023uet}. 
We produced 5000 realisations of this source model, by considering variations of the 
$\dNdS$ parameters inferred in~\cite{Amerio:2023uet}. In particular, we adopt the model related to the \verb|gll_iem_v07| foreground template \footnote{This is the latest foreground model provided by Fermi-LAT.} and $|b|<30^\circ$ Galactic plane cut, and we vary the $\dNdS$ within the estimated uncertainties by fitting a Gaussian Process (c.f.  \cite{Rasmussen2005}) to the $\dNdS$  output of the neural network \footnote{The reason for considering a Gaussian Process here is two-fold: to interpolate the $\dNdS$ to values of the flux not included in the neural net output, as well as to have an estimation of the uncertainty band of $\dNdS$ at such interpolated values. }. 
For each parameter configuration of the $\dNdS$, the $F_{\rm  iso}$ is consistently determined by the procedure proposed in~\cite{Amerio:2023uet}.
Note that the residual Poisson-distributed isotropic component $F_{\rm iso}$  is now lower than  $F_0$ in eq.~\eqref{eq:mapB}, since part of the flux $F_0$ is accounted for by the discrete sources. 

We note that both the $\mathcal{K}$ and $\mathcal{M}$ maps are simulated, i.e.~are ``theoretical skies'', not the ``measured sky''. The map ${\cal K}$ is deterministic, meaning we  simulate it by accounting for the Fermi PSF for the point-like sources whose fluxes and positions are given by the 4FGL catalogue, on the top of the assumed smooth background (the remaining contributions to the $\mathcal{K}$ map in eq. 2.3). In the case of the map ${\cal M}$, we rely on statistical information, in which the d$N$/d$S$ does not describe the position of any gamma-ray source, nor does it tell us their exact number. Rather, the integral of d$N$/d$S$ in a flux bin tells us the expected number of sources in that bin, while the actual number of sources is only defined up to Poisson fluctuations. Once such a number of sources per flux bin has been determined, we place them at uniformly random positions in the sky, accounting for the ${\cal S}$ term in Eq.~\eqref{eq:map}. In summary, we run the simulator developed in~\cite{Amerio:2023uet} in order to build the maps ${\cal M}$.

\section{Problem setting and statistical framework}
\label{sec:procedure}
Our first step is the determination on the null-hypothesis model, consisting solely 
of Poisson-distributed background emission.
To this end, we apply a mask of  $|b|<30^\circ$ around the Galactic plane, and remove the emission of all 
sources in the 4FGL-DR3 catalogue, further masking a $1^\circ$ disk around each source centroid. This cut roughly corresponds to the 90\% containement angle of the PSF and represents a good compromise between removing the bulk of the pointlike emission while avoiding a too large removal of the diffuse emission area. We checked that the results are not particularly sensitive to the exact cut used in the $1^\circ-2^\circ$ range. 
%
We then compute the expected photon counts $\lambda_i$ of model $\mathcal{B}$ (depending on $A_{\rm gal}$  and $ F_{0}$) in the pixel $i$  and, given the actual counts $k_i$, define the Poissonian likelihood  
\begin{equation}
{\cal L}_0=\prod_{i=1}^{N_{\rm pix}} \frac{\lambda_i^{k_i}e^{-{\lambda_i}}}{k_i!}\, .
\end{equation}
The adopted parameters $A_{\rm gal}$  and $ F_{0}$ are those maximising the likelihood above. For $N_{\rm side}=512$, we find $A_{\rm gal}=0.914$ and $F_0=4.81\cdot 10^{-7}$ cm$^{-2}$ s$^{-1}$ sr$^{-1}$.

\medskip
Our second step is to quantify the significance of additional point sources on top of the
null-hypothesis model.
For this purpose, we define a test statistic ($TS$) function in each pixel, quantifying how well the  observed photon counts agree with the background-only hypothesis:    
\begin{equation}
    TS_i \equiv \frac{(x_i - \lambda_i)^2}{\lambda_i}~,
\label{eq:TS}
\end{equation}
where we indicate with $\lambda_i$ the best-fit background-only model counts in the pixel $i$, computed according to eq.~\eqref{eq:mapB};  $x_i$ can be either the observed photon counts 
(when applied to the actual photon count map of data, leading to the single set $TS^{\rm data}_i$) or the  photon counts in pixel $i$ generated via synthetic maps. If the maps are generated according to  the hypothesis eq.~\eqref{eq:mapK}, the procedure yields the single set $TS^{\rm cat}_i$; if the maps are randomly taken from the model of eq.~\eqref{eq:map}, we obtain a number of sets $TS^{\rm sim}_i$ equal to the number of realisations.
The $TS$ choice of eq.~\eqref{eq:TS} is inspired by Pearson's $\chi^2$ test statistic for Poisson distributed counts,
whose variance $\sigma_i^2=\lambda_i$. 
This choice also loosely mimics the one adopted by the \Fermi-LAT collaboration, albeit no exact correspondence can be expected.  In fact, \Fermi-LAT uses a recursive procedure where, after a first step similar to the above, energy-dependent renormalisations of the background are allowed independently in each ``region of interest''.  We prefer to use a simple and coherent definition valid for all pixels in the sky, also because we do not assume a particular probability distribution for the $TS$ values. Rather, we use it as a ``signal interest label'', and derive a probabilistic interpretation for it from simulations of synthetic maps (see below). 

\begin{figure}[t]
\centering
\includegraphics[width=0.6\textwidth]{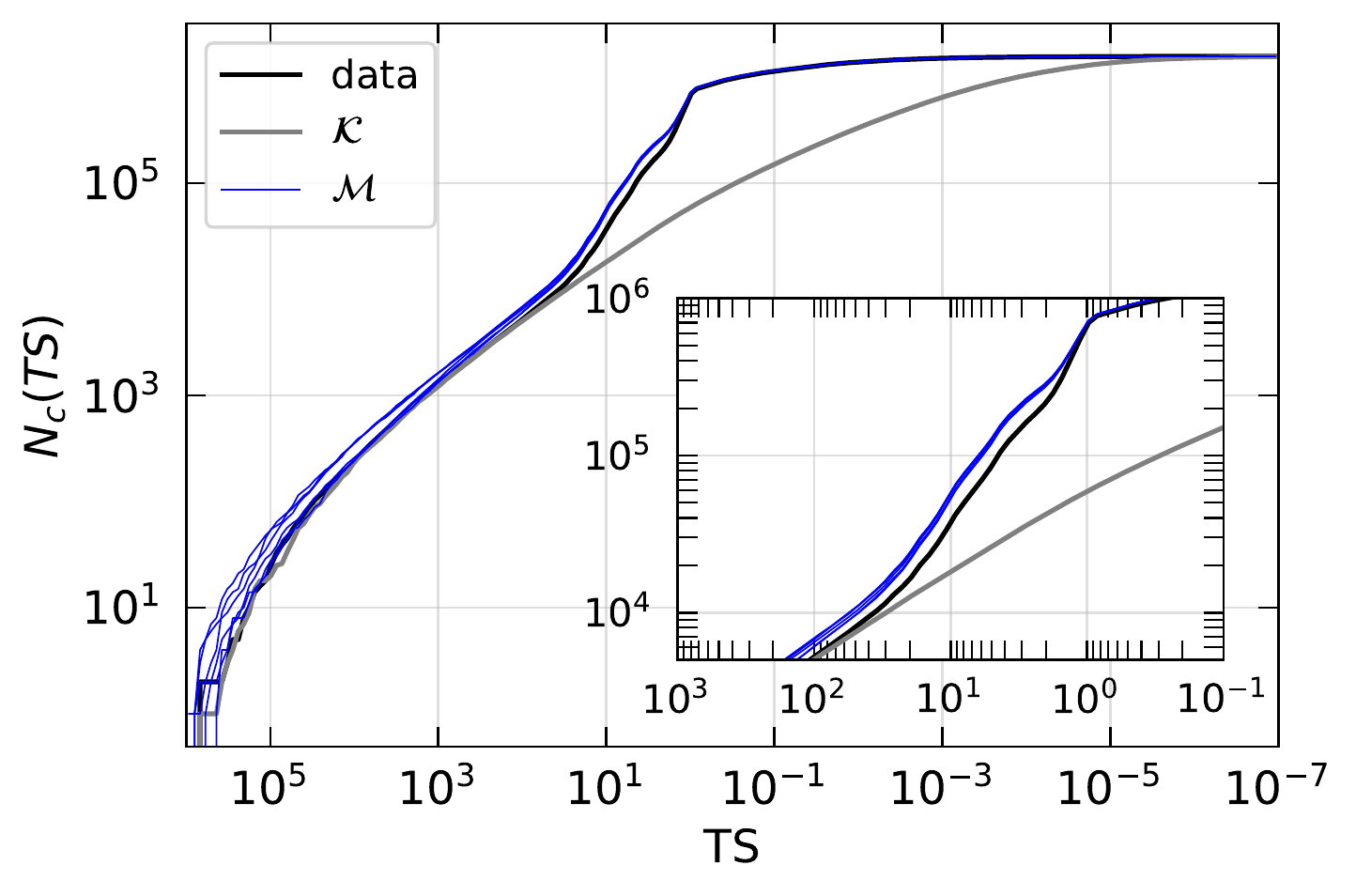}
\caption{Descending cumulative $TS$ distribution for the maps corresponding to: real data (black), 4FGL-DR3 catalogue (gray) and a few realisations of the $\dNdS$ source model (blue).
}
\label{fig:1}
\end{figure}

Obviously, since in our simulations the sources are uniformly distributed across the sky, the spatial distribution of $TS^{\rm sim}_i$  never coincides with the one obtained from the actual data, $TS^{\rm data}_i$, nor with the one from the 4FGL-DR3 catalogue, $TS^{\rm cat}_i$. However {\it the histogram of the $TS^{\rm sim}_i$ values should statistically match the actual one $TS^{\rm data}_i$, if the functional form and inferred parameters of the underlying $\dNdS$ are correct}.  
Since the determination of $\dNdS$ exploits information below the nominal catalogue's threshold, the statistical agreement between the actual map and the synthetic maps should hold at values below this threshold and inform us about new, faint sources. This is the gist of the argument that we exploit to extend {\it statistically} the \Fermi-LAT catalogue below threshold. This expectation is illustrated in figure~\ref{fig:1}, where we report the descending cumulative $TS$ distribution for real data (black), the 4FGL-DR3 catalogue source model, $\mathcal{K}$ (gray), and for several realisations of the $\dNdS$ source model, $\mathcal{M}$ (blue). While the distributions match rather well at high $TS$, the one associated to the catalogue departs from the data one well before the simulation's ones do. 

\medskip
As a next (and third) step, we attribute a statistical meaning to the $TS$ distributions. 
At large $TS$, we expect the observed map and the simulated ones to  agree well: Visible (but insignificant) discrepancies are expected at the highest values of $TS$, where Poisson fluctuations are important. At low $TS$ we expect the deviations from the two distributions to eventually become statistically significant. This is due to the interplay of the weakness of the sources and the systematic errors in the background template, that are more and more relevant at low fluxes. 

To assess how well the two distributions are in agreement, we apply a two-sample Kolmogorov-Smirnov (KS) test ~\cite{KS_test} to the normalised cumulative distribution functions of real data and simulated sky (or catalogue sky), both cut at a minimum $TS_*$, which we vary on a grid.
Their compatibility depends on the confidence level $1-\alpha$ at which we set the test. This is strictly speaking arbitrary, although conventionally $\alpha$ is set to 0.05 in most applications. We will actually consider three values of $\alpha$: 0.01, 0.05 and 0.1, to assess the dependence of the results on this meta-parameter. In general, increasing $\alpha$ yields more conservative constraints, as the criterion to reject the null hypothesis becomes more restrictive (i.e. the distributions should agree more closely). 

For the single catalogue model in eq.~(\ref{eq:mapK}), we present the results in terms of the $TS_*$ leading to  agreement with real data as a function of $\alpha$. 

In the case of the $\dNdS$ model, since we have multiple realisations, we quantify what is the fraction of simulations which passes the test against the true sky distribution as a function of $TS_*$ (and $\alpha$). This provides a {\it quality factor} $QF$ that we associate to the considered cut. Lacking a better criterion, a reasonable choice may be to require a majority of realisations to lead to an agreement (i.e. $QF>0.5$), but it is strictly speaking another meta-parameter which depends on the criteria of the user.  We will also show how results depend on $QF$, with $QF$'s closer to 1 associated to more restrictive requirements. Note that we will ``naively'' interpret the $QF$ as probability of the $TS$ being associated to a source, by considering the whole sample of simulations as equally viable (i.e. only the prior information based on the pixel statistics is used). In principle, one could refine this indicator by restricting the sample of simulations to the ones which reproduce the high-$TS$ distribution. We expect the actual probability to be somewhat higher than the one estimated via the $QF$. Also note that nowhere in our procedure we have relied on the 4FGL catalogue as input.~\footnote{Technically, we mask the catalogue position to fit the background model, but this is done only to lead to a better  background model and thus a more meaningful $TS$ scale. In principle, we could fit the background model without masking the catalogue source positions. That would alter our $TS$ scale, but the same bias will affect the simulated sky maps that we use to attribute a statistical meaning to the $TS$.}

The final step is how to identify the ``firing pixels'' in the true sky which are going to constitute the main deliverable of our work. For a fixed KS test significance $\alpha$ and a quality factor QF, we obtain a value $TS_*$ above which a fraction QF of simulations is compatible with the measured sky. Those pixels $i$ in the latter having $TS_i>TS_*$ will simply constitute of candidate pixels.

\medskip
We conclude this section with a general remark. We note that our procedure treats every pixel independently, assigning a $TS$ value to each of them, regardless of the $TS$ values of the neighboring pixels. In this respect, it is  useful to distinguish between a source and a ``firing pixel'': Multiple pixels may be firing (i.e.~having a sufficiently large $TS$) in association to a single source, depending on the choice of $N_{\rm pix}$, the intensity of the source and the PSF of the instrument (indirectly, thus, also on the spectrum of the source). This instance is more and more likely the smaller the pixel size, i.e.~the larger $N_{\rm pix}$, is. Conversely, multiple sources may be associated to a single pixel, notably for choices of $N_{\rm pix}$ which are too low. Clearly, for the exercise proposed here to be useful, we must require that the choice of $N_{\rm pix}$ leads to a number of pixels in the non-masked sky sufficiently larger than the current number of sources in the catalogue in the same region. At the same time, a pixel size  much smaller than the angular resolution would be meaningless. Keeping this in mind, for the sake of simulations, we fix $N_{\rm side}=1024$, which corresponds to an angular resolution of approximately $0.06^\circ$, in order to match the angular resolution of the Galactic foreground template. However, since the \Fermi-LAT angular resolution in the (1,10) GeV energy band is at best of the order of $0.15^\circ$ \footnote{\href{https://fermi.gsfc.nasa.gov/science/instruments/table1-1.html}{https://fermi.gsfc.nasa.gov/science/instruments/table1-1.html}}, for the sake of our pixel analysis, we will downgrade our simulations to $N_{\rm side}=512$, which corresponds to an angular resolution of approximately $0.12^\circ$. The choice of this value for $N_{\rm side}$ ensures that there are considerably more pixels than the number of resolved sources (approximately 500 times the number of resolved sources in the region $|b|>30^\circ$) and that the pixel size  of our analysis roughly matches the best resolution of the LAT at the energies of interest.
The $TS$ analysis is therefore performed with synthetic and real data maps pixelized with $N_{\rm side}=512$.

\section{Results}
\label{sec:catalogue}

We comment here on some summary statistics and synthetic analysis of our main results.
In table~\ref{tab:tsstar} we report, for different values of $\alpha$, the 
$TS_*$ above which the 4FGL-DR3 catalogue source model and real data pass the KS test (second column, $TS_*^{\rm cat}$), and the corresponding $TS_*$ where the $\dNdS$~simulated sky maps do the same ($TS_*^{\rm sim}$), at different values of $QF$.
\begin{table}[!h]
    \renewcommand{\arraystretch}{1.2}
    \centering
    \begin{tabular}{c|c|c|c|c}   
        $\alpha$ & $TS_*^{\rm cat}$ & $TS_*^{\rm sim}(QF=0.9)$& $TS_*^{\rm sim}(QF=0.8)$&  $TS_*^{\rm sim}(QF=0.5)$\\
        \hline 
       0.01 & 53 & 57 & 42 & 33\\ 
        \hline 
       0.05 & 62 & 147 & 48 & 36\\ 
        \hline 
       0.10 & 63 & 307 & 74 & 37\\
    \end{tabular}
\caption{Required $TS_*$ for three values of $\alpha$ (KS test confidence level) for the matching between 4FGL-DR3 catalogue source model and real data (second column, $TS_*^{\rm cat}$), and the corresponding $TS_*$ for the $\dNdS$~simulated sky maps vs real data (rest of columns, $TS_*^{\rm sim}$), at different values of $QF$.}  
\label{tab:tsstar}
\end{table}

\begin{table}[!h]
    \renewcommand{\arraystretch}{1.2}
    \centering
    \begin{tabular}{c|c|c|c}   
        $\alpha$ & $\widetilde{TS}_*^{\rm sim}(QF=0.9)$& $\widetilde{TS}_*^{\rm sim}(QF=0.8)$& $\widetilde{TS}_*^{\rm sim}(QF=0.5)$\\
        \hline 
       0.01  & 181 & 161 & 104 \\ 
        \hline 
       0.05  & 301 & 180 &  148\\ 
        \hline 
       0.10  & 400 & 200 &  160\\
    \end{tabular}
\caption{Required $TS_*$ for three values of $\alpha$ (KS test confidence level), for the $\dNdS$ simulated maps to match the 4FGL catalogue model map, at different values of $QF$. Note that the KS test is different from the one in table~\ref{tab:tsstar}, where the matching was with respect to the \Fermi-LAT data map.}  
\label{tab:ts4FGL}
\end{table}
The first conclusion is that, provided that the choice of $QF$ (and/or  of $\alpha$) is not too stringent, 
 our model maps $\mathcal{M}$ agree with the data better than the catalogue map does, i.e.~the agreement extends to lower $TS_*$\: Compare in particular the second with the last column in table~\ref{tab:tsstar}. This is just a manifestation of the fact that, at least for a majority of realisations, including sub-threshold sources provides a much better description of \Fermi-LAT sky map. On the other hand, requiring a too large $QF$ may be counter-productive: Even at high-$TS$, there is a non-negligible fraction of the simulations that does not agree with the data as measured by the KS test.  This manifests in the high values of $TS_*$ found in the $QF=0.9$ column and to some extent for $QF=0.8$ as well.

A second sanity check is that  the $\dNdS$~model is better in agreement with the data than with the catalogue: This manifests in the values of $\widetilde{TS}_*$ (see table \ref{tab:ts4FGL}) for which the $\dNdS$~model yields maps in agreement  with the catalogue map $\mathcal{K}$, hence indicated with a tilde. As we can see from the table, the $\widetilde{TS}_*$ values are significantly higher than the corresponding $TS_*$ reported in table~\ref{tab:tsstar}.

A further natural question to ask oneself is if the statistical procedure leads to a competitive sample of candidate source positions in the sky, when compared to the \Fermi-LAT catalogue analysis. After all, requiring an agreement of the $TS$ histograms in the KS-sense is a {\it qualitatively} different procedure than the one used by the collaboration to infer their catalogue. 
A direct comparison is impossible, if anything since our discretisation of the sky only allows us to identify firing pixels, which are not in a one-to-one correspondence with sources, as discussed in the previous section. However, for a reasonable choice of the pixel size as the one adopted here (see discussion at the end of section~\ref{sec:procedure}) we expect the number of  firing pixels to be a reasonable proxy of the number of underlying candidate sources. Our results are reported in table~\ref{tab:number_firing}.
\begin{table}[!h]
    \renewcommand{\arraystretch}{1.2}
    \centering
    \begin{tabular}{c|c|c|c|c}   
        $\alpha$ & $N^{\rm cat}$ & $N^{\rm sim}(QF=0.9)$& $N^{\rm sim}(QF=0.8)$& $N^{\rm sim}(QF=0.5)$\\
        \hline 
       0.01 & 7412 & 7050 & 8543  & 10068\\
        \hline 
       0.05 & 6707 & 3975 & 7891  & 9636\\
        \hline 
       0.10 & 6580 & 2492 & 5965  & 9409\\
    \end{tabular}
\caption{Number of firing pixels associated to the same cases of table~\ref{tab:tsstar}.}  
\label{tab:number_firing}
\end{table}

A comment on this result is that the number of pixels inferred from the catalogue map is in the ballpark of the sources reported in the \Fermi-LAT catalogue. This result was not obvious a priori. We may tentatively conclude that the power of the procedure we follow in identifying a firing pixel  is at least roughly comparable to the criterion the \Fermi-LAT collaboration follows in identifying a source. 
A second interesting conclusion is that the number of firing pixels can be significantly higher (up to a factor 1.5) if using the $\dNdS$~model maps, at least if sufficiently loose criteria are used for $QF$ and/or the level of KS agreement, i.e.~$\alpha$. We can conclude  that our procedure can be used to increase the number of candidate pixels in the sky that we can probabilistically attribute to sources, with respect to the ones one would deduce by merely using the existing catalogue.   

A question arises though: Are the identified pixels in the sky truly associated to sources? Of course we cannot answer this question without lengthy and careful follow-up studies and extensions of the present approach. We discuss some interesting directions in the next section. However, an interesting sanity check is that the bulk of the positions of \Fermi-LAT sources should match part of the recovered firing pixel positions. \begin{figure}[t]
\centering
\includegraphics[width=0.6\textwidth]{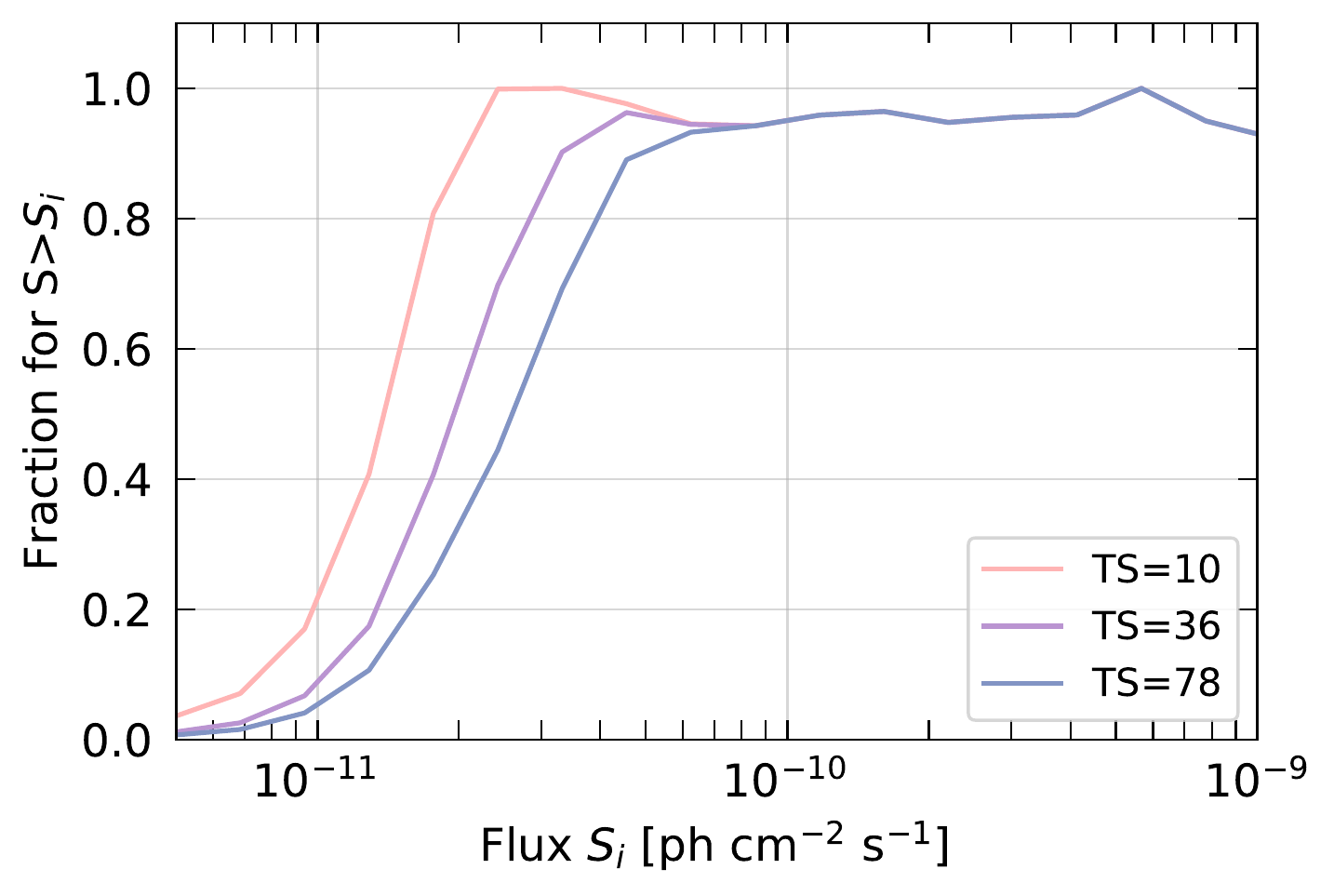}
\caption{Fraction of sources in the model $\mathcal{K}$ map, whose position corresponds to  a firing pixel for different values of $TS$. See text for details.}
\label{fig:2}
\end{figure}

In figure~\ref{fig:2} we compute, as a function of the flux, the fraction of all the pixels in the catalogue map ${\cal K}$, having a flux $S>S_i$, which coincide with the pixels of our  firing pixels map obeying the same flux condition, for different $TS$ cuts. The pixel coincidence is actually relaxed to a distance corresponding to the 68\% containment angle of the PSF, which naively accounts for the uncertainty of the quoted source centroids in the 4FGL catalogue.  
In figure~\ref{fig:2} we compute the fraction of all the pixels in the catalogue map ${\cal K}$ with flux $S>S_i$, whose positions are found in spatial coincidence with  pixels in our firing-pixels map obeying the same flux condition.\footnote{The pixel-wise spatial coincidence is actually relaxed to a distance corresponding to the 68\% containment angle of the PSF, which naively accounts for the uncertainty of the quoted source centroids in the 4FGL catalogue.} We display the fraction as a function of the flux and for different $TS$ cuts.

\begin{figure}[t]
\centering
\includegraphics[width=0.6\textwidth]{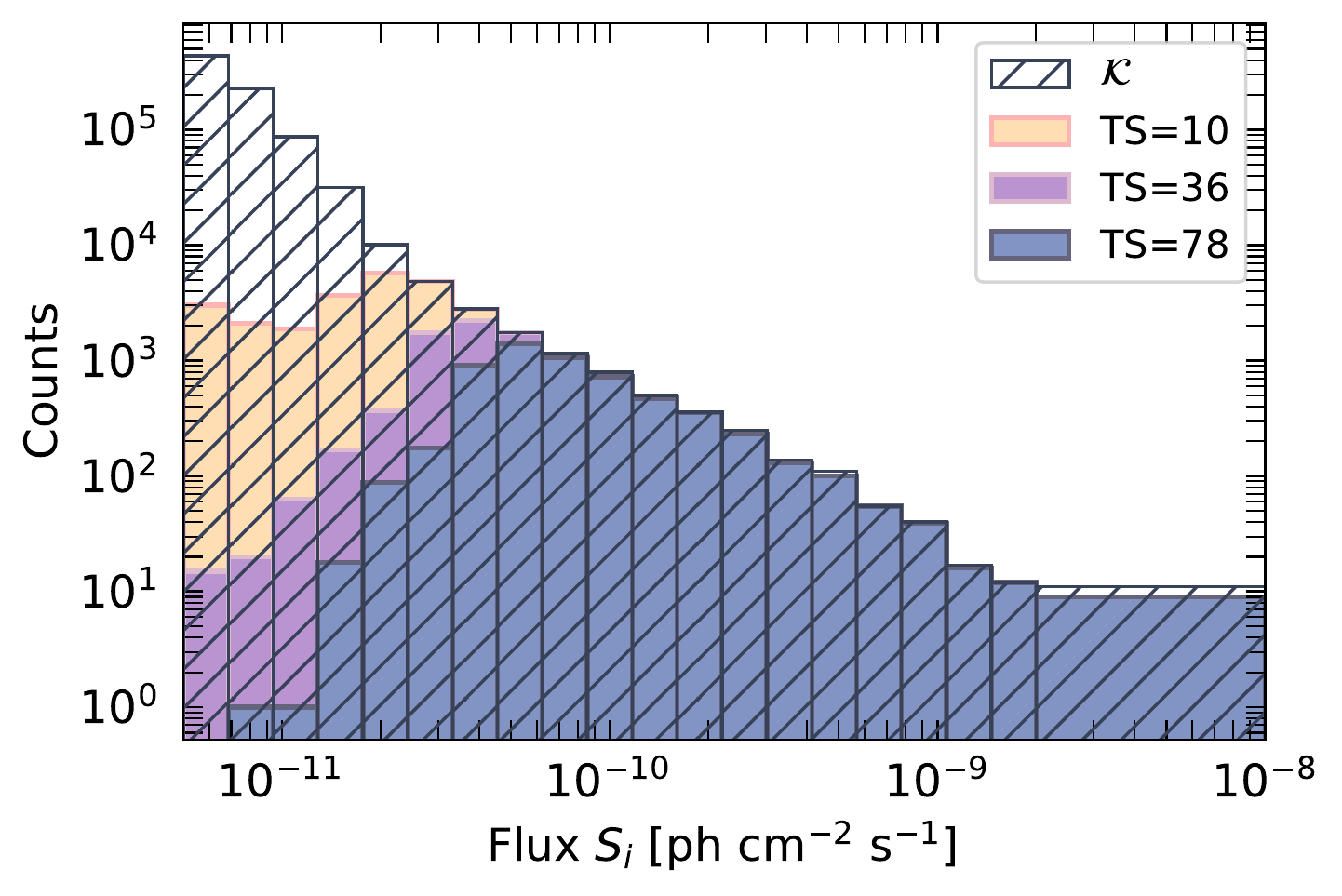}
\caption{Analogous to  figure~\ref{fig:2}, but for differential flux; i.e. number of pixels for a given flux bin. White histograms correspond to the whole catalogue map ${\cal K}$, while colored histograms are only those pixels in ${\cal K}$ coinciding (within the 68\% PSF containment angle) with the firing pixels of our \dNdS~model map. }
\label{fig:2b}
\end{figure}

Roughly~\footnote{More quantitative considerations require energy- and position-dependent assessments.}, \Fermi-LAT starts to loose sensitivity to sources below $2\times 10^{-10}\, {\rm cm}^{-2}{\rm s}^{-1}$~\cite{Zechlin:2015wdz,DiMauro:2017ing,Marcotulli:2020fpm,Amerio:2023uet}, above the rising region of the sigmoid-like curves of figure~\ref{fig:2}. As expected, modulo fluctuations, the fraction tends towards unity at high fluxes. Note that we do not expect perfect recovering of the sources in the \Fermi-LAT catalogue, anyway, either due to some simplifications (e.g.~we use energy-integrated quantities for our analysis, as opposed to the differential energy analysis by the LAT collaboration) or methodological differences, such as the pixelisation procedure or the iterative background estimate procedure in \Fermi. But it is clearly the case that we do recover the largest majority of the ``bona fide'' sources that we should. 

Figure~\ref{fig:2b} gives us a more detailed look at the information contained in figure~\ref{fig:2}. In this case we do not consider the ``cumulative'' number of pixels, but only those lying in a certain flux bin. The colored histograms correspond to the pixels in map ${\cal K}$ matching with the firing pixels of our \dNdS~model map (related to the numerator of the fraction computed in figure~\ref{fig:2}). The white histogram instead correspond to all the pixels in map ${\cal K}$ (related to the denominator of the aforementioned fraction). The sigmoid curve in figure~\ref{fig:2} grows close to unity at large $S$ since---barring fluctuations---the quasi-totality of sources in the catalogue are found within the adopted cuts. When lowering the cut $TS$, the peak of the colored histogram shifts to lower $S$, while still preferentially populating the high-$S$ tail (compare case $TS=36$ with case $TS=78$). If the $TS$ is lowered too much, as illustrated by the case $TS=10$, the matching starts developing a ``flat tail'' at lower $S$, hinting at a growing  random match component. Eventually, the curves in figure~\ref{fig:2} drop to zero at low $S$ not because of the scarcity of matchings (i.e., the numerator), but due to the faster growing number of pixels having flux above the indicated value (i.e. the denominator).

As a further check, we assess how  robust the selected catalogue is to variations of the background model. We compare the results obtained with the \verb|gll_iem_v07| template with the previous incarnation of the galactic foreground model developed by the Fermi collaboration: \verb|gll_iem_v05_rev1|. Since the $TS^*$ scale and related $QF$ are quantitatively different if the different backgrounds are used, we proceed by comparing the two results {\it at equal number of selected sources}.  We select $TS^*$ cuts delivering the same number of selected sources for the two cases and evaluate the exact overlap of pixel candidates. The results are reported in Tab.~\ref{tab:bck_comp}. As can be seen, within a few percent, the catalogues obtained agree.

\begin{table}[t]
    \renewcommand{\arraystretch}{1.2}
    \centering
    \begin{tabular}{|c|c|}  
    \hline
        N &  coincidence  \\
        \hline 
       3000 & 98.4\% \\
        \hline 
       5000 & 97.8\% \\
        \hline 
       9000 & 97.8\% \\
       \hline
    \end{tabular}
\caption{The fraction of firing pixels in common between the selections derived for the two galactic foreground models. The comparison is performed at equal number of selected sources for the two cases.}  
\label{tab:bck_comp}
\end{table}

The main output of our work consists of sky maps of ``firing'' pixels (i.e.~source candidates). These pixels have a significant $TS$ value among all pixels for which there is a statistical compatibility between simulations and real data, according to KS tests. The firing pixels depend on the significance $\alpha$ of the KS test, as well as on a ``quality factor'' $QF$, accounting for the fraction of simulated maps passing the KS test.
We provide our results electronically in the form of a Python package called \verb|gPCS|  (for gamma-ray Photon Count Statistics), as well as a summary \verb|FITS|, both available on Zenodo: \href{https://doi.org/10.5281/zenodo.8070852}{https://doi.org/10.5281/zenodo.8070852}. The Python package can be used to determine the firing pixels given either a $TS_*$ or a quality factor and $\alpha$ parameter. The Python package also includes a convenience function to export the firing pixels in the form of a \verb|FITS| file, and we provide an example on how to compute the summary table provided with the package.
\\
The \verb|gPCS| Python package can be easily installed using \verb|pip|, with the command:
\begin{lstlisting}[language=iPython]
pip install gPCS
\end{lstlisting}
In order to get the firing pixels, given either $TS_*$ or a quality factor and $\alpha$, we can use the following snippet of code:

\begin{lstlisting}[language=iPython]
import numpy as np
from gPCS import gPCS

# specify manually a TS_star
TS_star = 36
pixel_firing = gPCS.get_firing_pixels(TS_star, filter=False)
print(len(pixel_firing))

# Compute the TS_star from a chosen QF and alpha
# [alpha must be 0.01, 0.05 or 0.1]
QF = 0.5
alpha = 0.05
TS_star = gPCS.get_TS_from_QF(QF, alpha=alpha)
pixel_firing = gPCS.get_firing_pixels(TS_star, filter=False)
print(len(pixel_firing))
\end{lstlisting}
It is possible to export a \verb|FITS| table using the \verb|export_fits_table| function. For example to reproduce the table provided with the library we can use:

\begin{lstlisting}[language=iPython]
gPCS.export_fits_table(filename="firing_pixels.fits", QF=0.50, alpha=[0.01, 0.05, 0.1])
\end{lstlisting}
The \verb|FITS| table has the following fields:
\begin{lstlisting}[basicstyle=\ttfamily]
    pixel   : pixel index
    TS      : TS value for each pixel
    QF_best : QF value obtained by considering all the simulations
    QF_min  : lower bound of the QF range
    QF_max  : upper bound of the QF range
\end{lstlisting}
For further details on the package usage, please refer to the documentation and examples: \\
\href{https://github.com/aurelio-amerio/gPCS}{https://github.com/aurelio-amerio/gPCS}.

\section{Discussion and conclusions}
\label{sec:conclusions}

We have presented an approach to statistically push the \Fermi-LAT sensitivity to point-like sources at high latitudes below  the current threshold for detection, leveraging on the fact that the underlying source count  distribution function is constrained even for lower test statistics ($TS$). This function has been recently re-derived using machine learning methods, yielding results in  accordance with existing pixel statistics analyses. This information, not accounted for in the current catalogue construction procedure, allowed us to provide a catalogue of directions (i.e.~firing pixels) in the sky likely to be associated to sources, albeit only in a probabilistic sense, that we assessed via a Kolmogorov-Smirnov (KS) test. The actual catalogue of directions depends on meta-parameters such as the confidence level $1-\alpha$ of the KS test, or the fraction $QF$ of simulated skies for which the simulation agrees with data above a certain $TS$ level. For reasonable choices of these meta-parameters, we found a number of ``firing pixels'' $\sim$50\%  higher than what one would infer from a catalogue only including sources listed in the latest incarnation of the \Fermi-LAT catalogue. Our results also pass some sanity checks, like the fact that \Fermi-LAT catalogue sources which are luminous enough are found within the directions inferred by our method, within the angular resolution of the instrument. If compared with the ``local'' iterative procedure followed by \Fermi-LAT to establish their catalogue, another possible advantage of our catalogue of directions in the sky is that its $TS$ scale has a homogeneous meaning, and may be more suitable for global statistical analyses.  

We provide our results in digital form, allowing the user to select the meta-parameters $QF$ and $\alpha$ to be more or less restrictive in their selection criteria, hopefully adapting to a wide range of applications. The most obvious one we can think of consists of cross-correlation studies, either multi-wavelength or multi-messenger, where the statistical advantage of  a significantly larger sample more than compensates having a (sub-leading) fraction of spurious directions. 
Another straightforward use of our results could be a guided source search and identification program via multi-wavelength studies, which would also help transforming some of these candidate source directions into  {\it bona fide} sources. In turn, these studies may further benefit of machine learning techniques that have been recently proposed to ease the  probabilistic classification of unassociated sources~\cite{Bhat:2021wtb}.

We stress that, although our results are based on the source-count distribution as inferred from a specific machine learning analysis of the photon counts, the method developed here to identify ``firing pixels'' can be applied to any other determination of the $\dNdS$ of the high-latitude sky, and it is, in this respect, independent on the choice of the $\dNdS$ as long as the latter correctly describes point sources in the faint regime.

Since our work had primarily a methodological proof-of-principle motivation, we have resorted to a couple of technical simplifications: We worked with an energy-integrated spectrum and we limited ourselves to high Galactic latitudes. We could rather naturally lift the former approximation  by a brute force approach. We expect the main complication to be computational, but the problem should remain affordable in a reasonable time. Also the latter limitation could be in principle lifted, extending the analysis to low Galactic latitudes. The main problem we can anticipate is that the existing background models fail to perform as well at low latitudes as they do at high latitudes. One also needs to include multiple populations contributing to both resolved and unresolved emission, see e.g.~\cite{Calore:2014oga} and sec. 6.2 in ~\cite{Fermi-LAT:2015bhf} for the role played by Galactic millisecond pulsars in this context. In the worst case, the method may not allow one to gain much more with respect to more traditional techniques. Depending on the outcome, one may however think of multi-zone models for the background or the like.  

A third direction for future progress is to attach a  more realistic probability scale to the $TS$ maps. We have constructed our synthetic skies only based on the statistical properties for the sources previously deduced via the pixel analysis. However, only a subset of these simulations is consistent with the properties of the actual sky: Notably at high flux where only a few sources exist, the associated $TS$ distribution significantly depends on the distribution of the sources in the sky randomly falling in a lower or higher background region (for an illustration, see the large variability at high $TS$ in figure~\ref{fig:1}). As briefly discussed in the article, a promising approach would be to narrow the sample of admissible synthetic skies used to compute $QF$ to the sole simulations which are statistically consistent with the $TS$ pattern of the most luminous sources. Such `constrained' simulations should then improve the reliability of the $QF$ scale. Another working direction with the same goal could be to improve the description of background and/or the parameterization for sub-threshold sources in the pixel analyses beyond what considered in~\cite{Amerio:2023uet}. This may bring the simulation distribution illustrated in figure~\ref{fig:1} in even better agreement with the $TS$ histogram. Finally, exploring alternative  metrics to the agreement in the KS sense is also a direction potentially worth exploring. 

\begin{acknowledgments}
We would like to thank Ángela Fernández-Pascual and Carlos M. Alaíz-Gudín for their valuable collaboration in the first stages of this project.  We thank S. Manconi for careful reading of the paper and her comments.
F.C.~and P.D.S.~acknowledge support by the ``Agence Nationale de la Recherche'', grant n. ANR-19-CE31-0005-01 (PI: F.~Calore). A.A. and B.Z. acknowledge the support from Generalitat Valenciana through the  ``GenT program'', ref.: CIDEGENT/2020/055 (PI: B.~Zaldivar). This research was partially supported by the computing infrastructure
``Artemisa'', funded by the European Union ERDF and
Comunitat Valenciana, as well as the technical support
provided by the Instituto de Física Corpuscular, IFIC
(CSIC-UV).

\end{acknowledgments}

\bibliography{biblio}

\end{document}